\documentclass{lhep}

\journal{}
\vol{}
\jyear{}
\pages{} 

\received{}
\published{}

\begin{document}

\title{On the question of the analysis of $J/\psi \rightarrow \rho\pi \rightarrow \pi^+\pi^-\pi^0$}
\author{K.Yu.~Todyshev,\auno{1,2}}
\address{$^1$ Budker Institute of Nuclear Physics, 11, akademika
 Lavrentieva prospect,  Novosibirsk, 630090, Russia}
\address{$^2$ Novosibirsk State University, 1, Pirogova street,  Novosibirsk, 630090, Russia}
\begin{abstract}
This paper presents a method for the analysis of process
$J/\psi \rightarrow \rho\pi \rightarrow \pi^+\pi^-\pi^0$ 
based on the consideration of the angles of expansion of finite pion pairs.
The proposed approach makes it possible to effectively carry out
selection of events in both neutral and charge-conjugate modes 
of the decay of $J/\psi \rightarrow \rho\pi$.
Application of the method for the analysis of similar three-body decays
in some cases will simplify the analysis and refine current results.
\end{abstract}


\maketitle

\section{Introduction}
\label{sec:intro}
The interest in the issue of effective separation of the  $J/\psi
\rightarrow \rho\pi$ decay modes arose from considering the
experimental results of the analysis of the $J/\psi$ decay into three
pions. The main contribution to this process is determined by
the decay of $J/\psi\rightarrow \rho \pi$ followed by the decay 
$\rho$ meson into two pions. The probability of this process
was measured in a number of experiments
\cite{MRK1:1976,CNTR:1976,DASP:1978,PLUT:1978,MRK2:1983,MRK3:1988,BES:1996,BES:2004,BABAR:2004}
and is the highest among $J/\psi$ decays with an intermediate hadron
resonance $\mathcal{B}_{\rho\pi}(J/\psi)=1.69 \pm 0.15\%$ \cite{PDG:2022}.
In addition, the ratio of partial widths
$\Gamma_{\rho^0\pi^0}/\Gamma_{\rho\pi}$ was measured in 
\cite{MRK1:1976,CNTR:1976,DASP:1978,PLUT:1978,MRK3:1988}, whence,
taking into account the result of the latest 
experiment \cite{MRK3:1988}, the PDG \cite{PDG:2022} gives the value
$\mathcal{B}_{\rho^0\pi^0}(J/\psi)=(5.6 \pm 0.7)\times 10^{-3}$.
A feature of the available experimental measurements of the
$\mathcal{B}_{\rho\pi}(J/\psi)$ value is a significant discrepancy
between the results of early experiments
\cite{MRK1:1976,CNTR:1976,DASP:1978,PLUT:1978, MRK2:1983,MRK3:1988,BES:1996} and later measurements by the
collaborations BES \cite{BES:2004} and BABAR \cite{BABAR:2004}.
Resulting error scaling factor, according to PDG\cite{PDG:2022} is 2.4.
These reasons make it interesting to continue the study
both the $J/\psi\rightarrow \rho \pi$ process and the entire set of
$J/\psi$ decay processes leading to a three-pion final state.

\begin{figure}[h!]
\centering
\centering\includegraphics[width=0.45\textwidth]{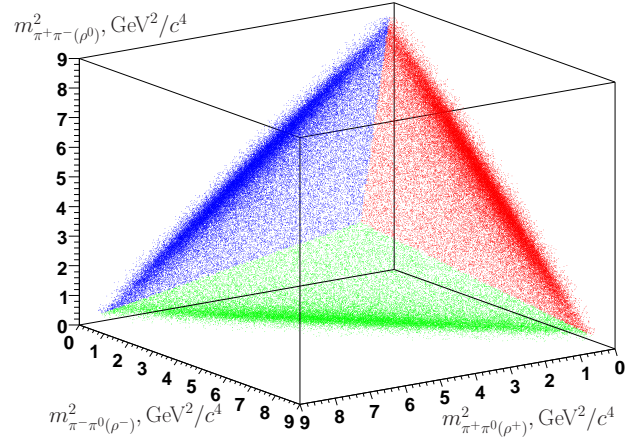}
\caption{The distribution of events over the squares of the invariant masses of
pion pairs  $m^2_{\pi^+\pi^0},m^2_{\pi^- \pi^0},m^2_{\pi^+\pi^-}$
obtained by MC simulation of the process  $J/\psi\rightarrow\rho
\pi$. The simulation was performed within the framework of the KEDR
experiment software \cite{KEDR:2013}. The red, blue and green regions
correspond to the conditions that select the $J/\psi$ decay modes
$\rho^+ \pi^-$, $\rho^-\pi^+$ and $\rho^0 \pi^0$, respectively, as
described in the text of the article.
\label{fig1}
}
\end{figure}

The distribution of squared invariant masses of pion pairs $\pi^+ \pi^0$, $\pi^- \pi^0$, $\pi^+ \pi^-$ is shown in three dimensions in the figure
\ref{fig1}. Any of the three 2D projections of a 3D distribution is a
Dalitz plot \cite{Dalitz:1954}. Plots of this kind are widely
used in the analysis of various processes, including those for the
three-pion $J/\psi$ meson decay.
Complexity of measuring the partial probabilities of modes $ J/\psi \rightarrow \rho^+ \pi^-$, $J/\psi \rightarrow \rho^-\pi^+$, $J/\psi \rightarrow \rho^0 \pi^0$ in such an approach is associated with a significant overlap of distributions of squared invariant masses
of pions pairs, as well as the need for a reliable description of the resolution function for a two-dimensional distribution over invariant masses.
The inaccuracies of this function, as a rule, introduce a noticeable
systematic uncertainty into the result.

\section{The main idea of analysis of the process $J/\psi\rightarrow\rho \pi$}
\label{sec:idea}
As an alternative to the two-dimensional fitting of Dalitz plots, one
can propose the following analysis procedure. 
At the initial stage, three subsets of events are selected in
accordance with the following conditions:
$\cos{\theta_{\pi^+\pi^0}}\!\!>\!\!\cos{\theta_{\pi^+\pi^-}} \wedge
\cos{\theta_{\pi^+\pi^0}}\!\!>\!\!\cos{\theta_{\pi^-\pi^0}}$,
$\cos{\theta_{\pi^-\pi^0}}\!\!>\!\!\cos{\theta_{\pi^+\pi^-}} \wedge
\cos{\theta_{\pi^-\pi^0}}\!\!>\!\!\cos{\theta_{\pi^+\pi^0}}$,
 and
$\cos{\theta_{\pi^+\pi^-}}\!\!>\!\!\cos{\theta_{\pi^-\pi^0}} \wedge
\cos{\theta_{\pi^+\pi^-}}\!\!>\!\!\cos{\theta_{\pi^+\pi^0}}$. 
Here and further, $\theta_{\pi^+\pi^0}$, $\theta_{\pi^+\pi^-}$ and
$\theta_{\pi^-\pi^+}$ are the angles between the momentum vectors of the
corresponding $\pi$ mesons.

\begin{figure}[h!]
\centering
\includegraphics[width=0.45\textwidth]{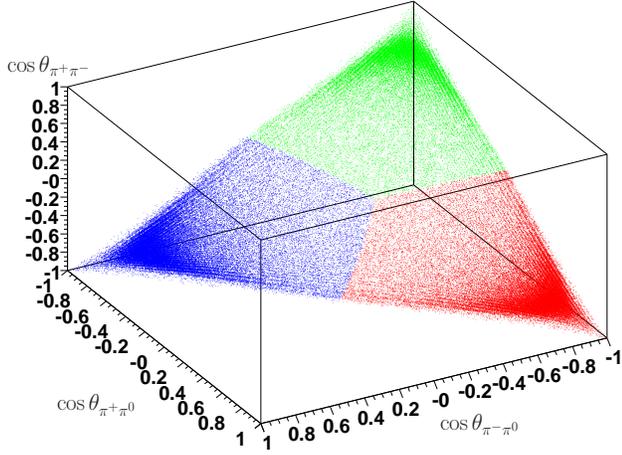}
\caption{Simulation data of the process $J/\psi\rightarrow\rho\pi$. The
distribution of events over the cosines of the angles
$\cos{\theta_{\pi^+\pi^0}},\cos{\theta_{\pi^-\pi^0}},\cos{\theta_{\pi^+\pi^-}}$
is given. The red, blue, and green regions correspond to the
conditions that single out, respectively, the $J/\psi$ meson decay
modes $\rho^+\pi^-$, $\rho^-\pi^+$ and $\rho^0\pi^0$.
\label{fig2}}
\end{figure}

Figure \ref{fig2} shows the distribution of events over the values of the
cosines of the angles of expansion of pion pairs in a
three-dimensional form. The criteria listed above determine the
boundaries of the regions of three subsets of events, where one or
another mode of the process under study  predominates.

The next stage of the analysis consists in constructing the
distributions of the invariant mass of the positively and negatively
charged $\rho$ meson according to the first
and the second subset of events, as well as the distribution of the
invariant mass $\rho^0$ meson over the events of the third subset.
The simultaneous fitting of the three resulting distributions involves the
calculation of all necessary parameters, as in the case of fitting
two-dimensional Dalitz plot. Let us discuss the advantages of the
proposed approach.

Consider the events corresponding to the selection conditions
$\cos{\theta_{\pi^+\pi^-}}<\cos{\theta_{\pi^-\pi^0}} \vee \cos{\theta_{\pi^+\pi^-}}<\cos{\theta_{\pi^+\pi^0}}$.
These criteria allow to reject most of the $J/\psi \rightarrow
\rho^0\pi^0$ events, which makes it possible to single out 
the $J/\psi \rightarrow \rho^+ \pi^-$ and $ J/\psi \rightarrow
\rho^-\pi^+$ ''conditional'' in its purest form.

\begin{figure}[h!]
\centering
\includegraphics[width=0.5\textwidth]{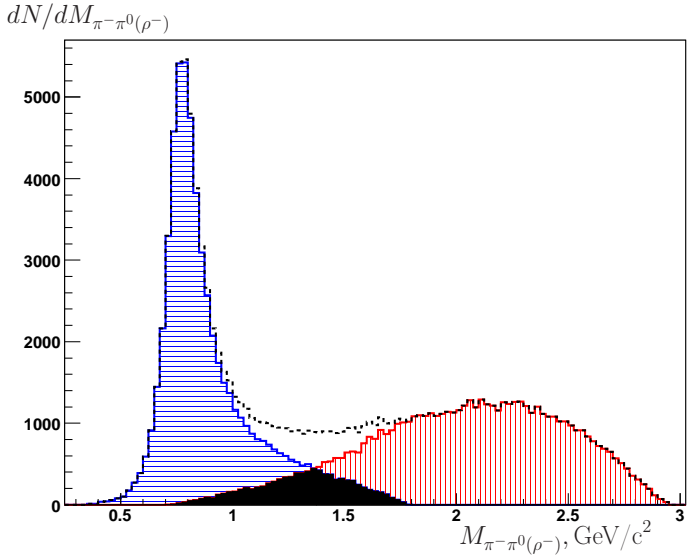}
\caption{ The invariant mass distribution of the pair $\pi^-$ and $\pi^0$ under the conditions $\cos{\theta_{\pi^+\pi^-}}<\cos{\theta_{\pi ^-\pi ^0}}\vee\cos{\theta_{\pi^+\pi^-}}<\cos{\theta_{\pi^+\pi^0}}$.
\label{fig3}}        
\end{figure}

Figure \ref{fig3} shows the distribution of the invariant mass for
the  $\pi^-\pi^0$ pair for events that meet the above criteria.  The resulting histogram is marked with a dotted line.
The area with horizontal blue hatching meets the condition
$\cos{\theta_{\pi^-\pi^0}}\!>\!\cos{\theta_{\pi^+\pi^0}}$, and the
part shaded with vertical red lines is the inverse relation
$\cos{\theta_{\pi^-\pi^0}}\!<\!\cos{\theta_{\pi^+\pi^0}}$.
The area of the intersection of the distributions determined 
by this condition is approximately $7.4\%$ of all events in the histogram.

\begin{figure}[h!]
\centering
\includegraphics[width=0.5\textwidth]{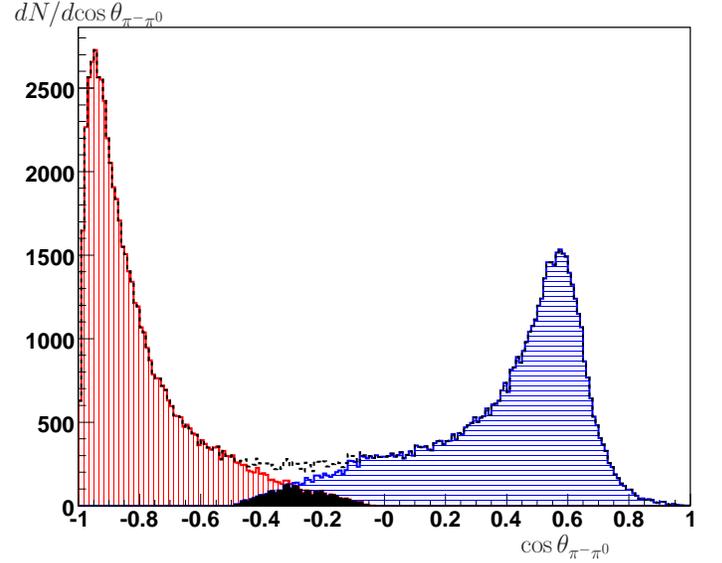}
\caption{\label{fig4} Distributions of the cosine of the angle between
  $\pi^-$ and $\pi^0$ under the conditions 
  $\cos{\theta_{\pi^+\pi^-}}<\cos{\theta_{\pi^-\pi^0}} \vee \cos{\theta_{\pi^+\pi^-}}<\cos{\theta_{\pi^+\pi^0}}$.
}        
\end{figure}

Consider now the distribution of the cosine of the angle between the
momentum vectors of $\pi^-$ and $\pi^0$ under the same conditions in the
figure \ref{fig4}.  The overlap area of the event subset for one-dimensional distributions in magnitude
$\cos{\theta_{\pi^-\pi^0}}$ in this case is $2.3\%$. 
The resulting value characterizes the overlap of subsets of events $J/\psi
\rightarrow \rho^+ \pi^-$ and $J/\psi \rightarrow \rho^- \pi^+$ on the
three-dimensional plot  of the angles of expansion of $\pi$ meson
pairs.

A similar parameter at the boundaries of the intersection of charged modes with a subset of events
neutral mode is around $2.5\%$. In all possible variants of
intersections, approximately a threefold advantage remains compared to
the values of overlapping distributions of invariant masses.
 
The event compaction effect of the ``corners'' area in figure \ref{fig2} is a
consequence of a simple fact. The shell or boundary of the area
occupied by the Dalitz plot for a three-pion decay corresponds to
"collinear" events, when the momenta of two particles are directed
against the direction of motion of the third particle. In this case,
the events of each of the $J/\psi \rightarrow\rho \pi$ decay modes
are located in the region close to the corresponding side of the
triangle in the figure \ref{fig1}, and on the cosine plot of the
expansion angles of $\pi$ meson pairs in the corresponding part of
the 3D drawing in figure \ref{fig2}.

It is possible to find a relation between the observed angle of
expansion of a pair of charged pions and their invariant mass, 
in an approximation where pion masses can be neglected. 
In this case, $2E_1E_2(1-\cos{\theta_{12}})=m_{12}^2$, where $E_1$ and $E_2$ 
are energies of the corresponding pions. Consider the 
$\pi^{-}\pi^{0}$ pair. Under the condition $\cos{\theta_{\pi^-\pi^0}}\!\!>\!\!\cos{\theta_{\pi^+\pi^-}} \wedge
\cos{\theta_{\pi^-\pi^0}}\!\!>\!\!\cos{\theta_{\pi^+\pi^0}}$, the
observable value is $\cos{\theta_{\pi^-\pi^0}}>-0.5$
(Fig. \ref{fig4}), from here one can determine that $m^2_{12}< 3 E_1
E_2$. The maximum of the product $E_1 E_2$ is reached at $E_1 = E_2= \frac{s-m_\pi^2+m_{12}^2}{4\sqrt{s}}$,
hence $m^2_{12}< \frac{3 s}{10}$, holding the next order of smallness
in pion mass leads to $m^2_{12}< \frac{8}{5}(\frac{3s
}{16}+4m_\pi^2)$. In numerical form, this corresponds to the condition
$ m_{12} < 1.73$~GeV$/c^2$ (for $s=M_{J/\psi}^2$).

\begin{figure}[h!]
\centering
\includegraphics[width=0.45\textwidth]{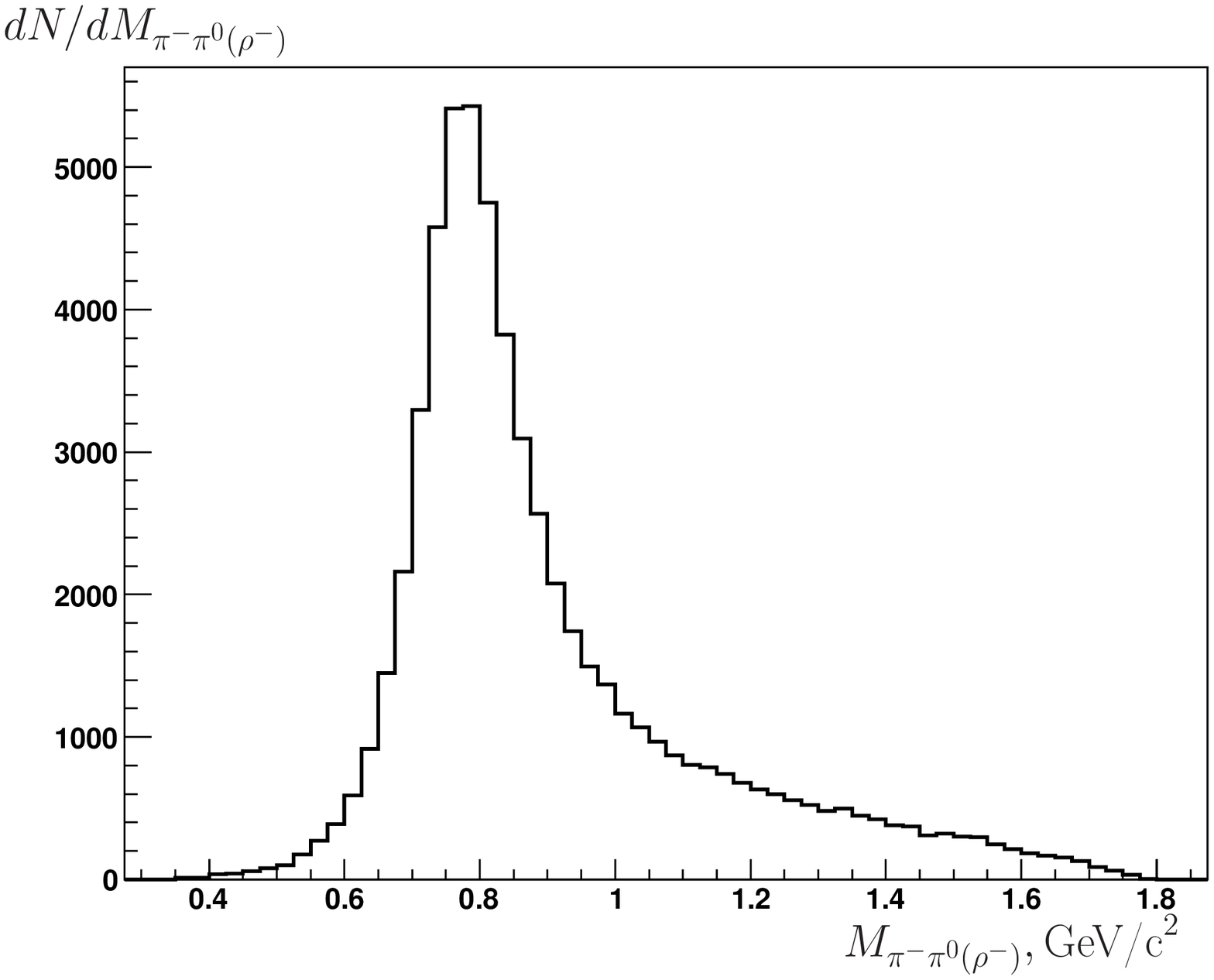}
\caption{\label{fig5}
Distribution of the invariant mass of the pair $\pi^-\pi^0$  under
the conditions  $\cos{\theta_{\pi^-\pi^0}}\!\!>\!\!\cos{\theta_{\pi^+\pi^-}} \!\wedge\! \cos{\theta_{\pi^-\pi^0}}\!>\!\cos{\theta_{\pi^+\pi^0}}$.}
\end{figure}

\begin{figure}[h!]
\centering
\includegraphics[width=0.45\textwidth]{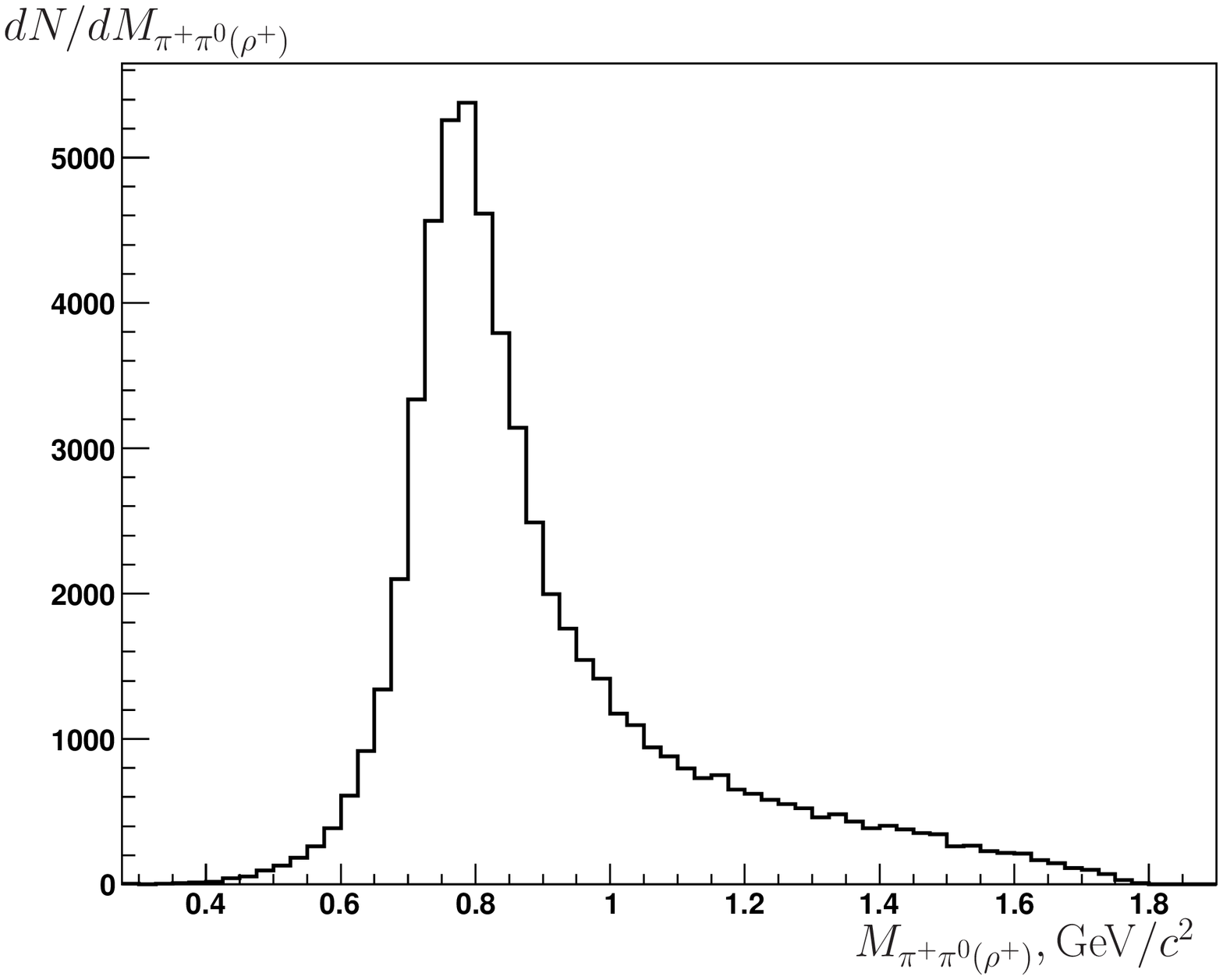}
\caption{\label{fig6}
Distribution of the invariant mass of the pair $\pi^+\pi^0$   under
the conditions $\cos{\theta_{\pi^+\pi^0}}\!\!>\!\!\cos{\theta_{\pi^+\pi^-}}\!\wedge\!\cos{\theta_{\pi^+\pi^0}}\!>\!\cos{\theta_{\pi^-\pi^0}}$.
}
\end{figure}

\begin{figure}[h!]
\centering
\includegraphics[width=0.45\textwidth]{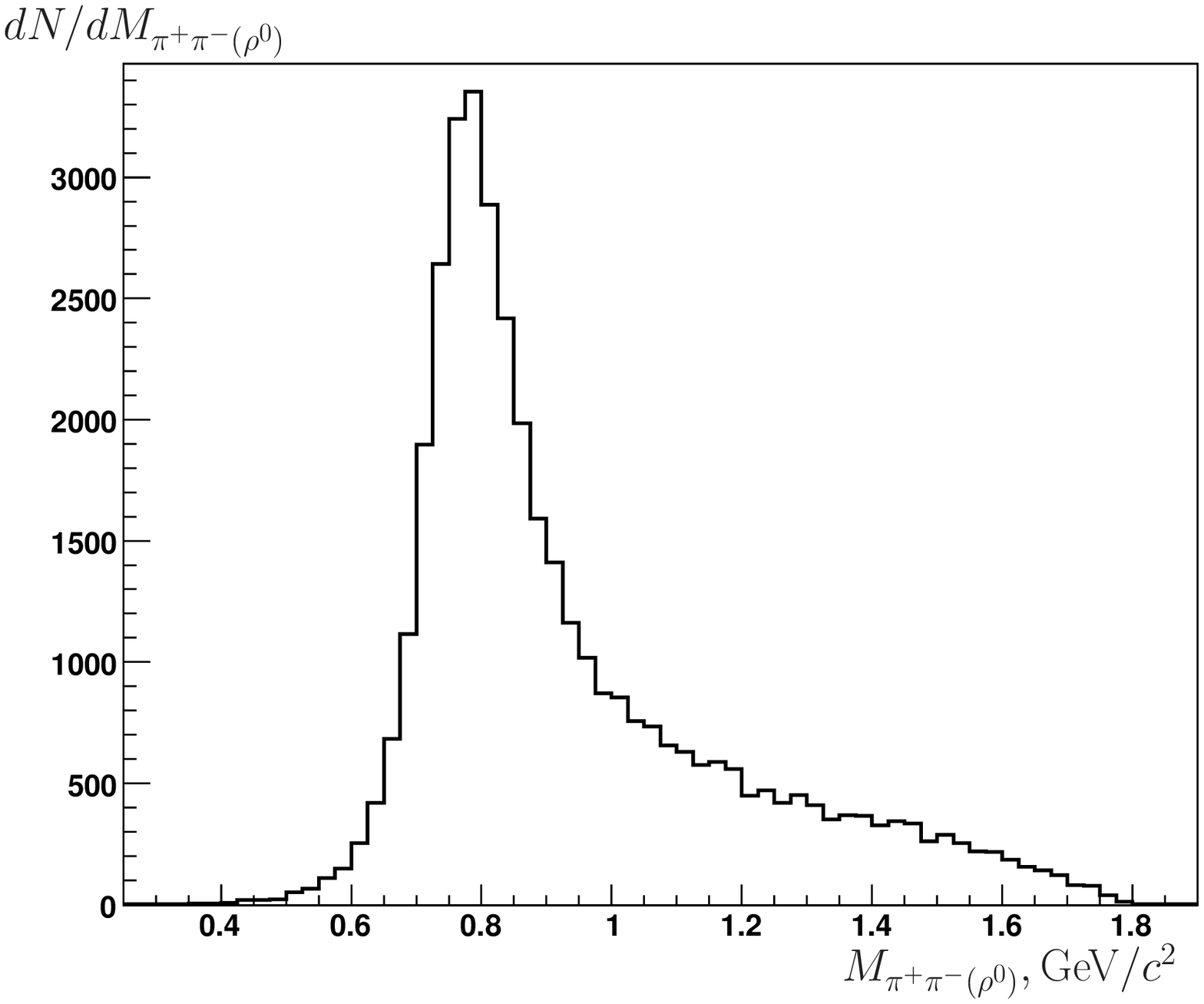}
\caption{\label{fig7} Distribution of the invariant mass of the pair $\pi^+\pi^-$  under
the conditions $\cos{\theta_{\pi^+\pi^-}}\!\!>\!\!\cos{\theta_{\pi^-\pi^0}} \wedge \cos{\theta_{\pi^+\pi^-}}\!\!>\cos{\theta_{\pi^+\pi^0}}$.}
\end{figure}

Using the above conditions on the cosines of the pion pair expansion
angles, it is possible to represent the entire set of events in the
form of one-dimensional distributions of invariant masses for pion
pairs, which is demonstrated in the figures \ref{fig5}, \ref{fig6} and
\ref{fig7}. The upper bound of each of the distributions agrees well
with the constraint found above.

\begin{figure}[h!]
\centering
\includegraphics[width=0.45\textwidth]{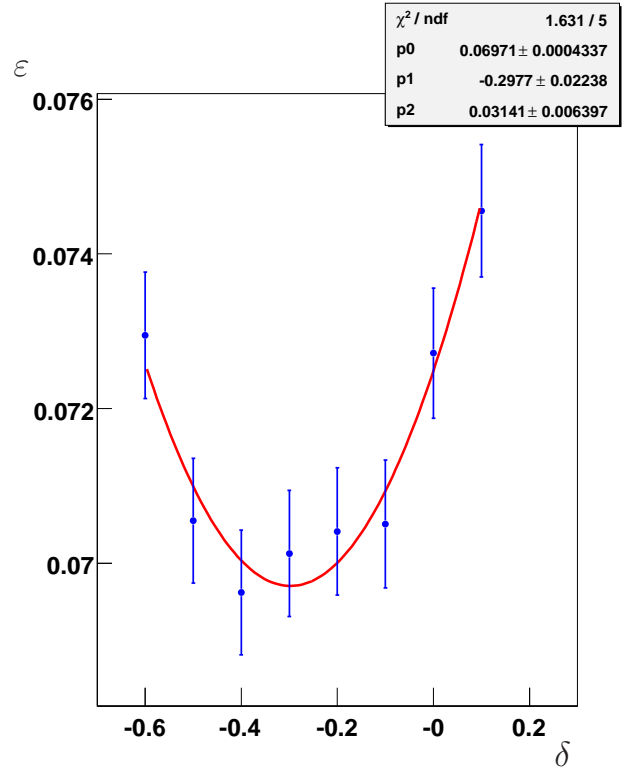}

\caption{\label{fig8}  Dependence of the sum of overlaps of subsets of
distinguished events $\varepsilon$ on the value $\delta$, which 
sets the shift to the value $\cos{\theta_{\pi^+\pi^-}}$ under the
conditions that determine the boundaries of the regions of the
$J/\psi$ decay modes under study.
}
\end{figure}

Consider the problem of optimizing the mode factorization procedure
for the process $J/\psi\rightarrow\rho\pi$, that is, we determine the possible region of
each of the modes so that the sum of the overlaps of the subsets
$\varepsilon$ is minimal.
Introduce the parameter $\delta$ and change
"$\cos{\theta_{\pi^+\pi^-}}$" to
"$\cos{\theta_{\pi^+\pi^-}}+\delta$" in all conditions listed
above.  Positive values of $\delta$ correspond to an increase in the
region where  the events of the $J/\psi\rightarrow\rho^0\pi^0$ mode
are distinguished, and negative values correspond to its
decrease. Figure \ref{fig8} shows the resulting dependence.

The optimal value of $\delta$ is found to be equal to $-0.3$. At the same time, 
the region of the minimum of the chosen optimization function 
is rather wide, and its difference from the value at zero $\delta$ 
is insignificant. Modification of the criteria on
$\cos{\theta_{\pi^+\pi^0}}$ and $\cos{\theta_{\pi^-\pi^0}}$ by
introducing additional offsets is not required due to the symmetry
selection conditions for these parameters, which is confirmed by
numerical calculation.

The observed asymmetry of the optimal boundaries of 
the selected events is associated with a lower efficiency of
registration  of the $J/\psi\rightarrow\rho^0\pi^0$ 
process compared to charged modes. When analyzing real experimental
data, the optimal  criteria for mode factorization may not correspond
exactly to the conditions described in this article due to the features of the
detector, the influence of background events, and the interference of 
the decay under study with other processes. 
Nevertheless,  selection criteria for the angles of expansion 
of  pion pairs of registered $\pi$ mesons in the process 
$J/\psi  \rightarrow \rho\pi$ significantly increase the efficiency of
 separation of various modes of a given decay. 
Such criteria are more efficient than restrictions on the squares of
the invariant masses and the invariant masses themselves for
processes  of this kind, if only one-dimensional distributions are
considered. Factorization of different $J/\psi \rightarrow \rho\pi$
decay modes can be useful for a more accurate measurement of the
partial probability this process.

\section{Conclusion}
\label{sec:conc}
A number of critical remarks should be made. 
The problem of interference of the main $\rho \pi$ channel 
with the $J/\psi \rightarrow \rho(1450) \pi$ decay and other 
possible processes leading to the three-pion state is not 
considered here, but the corresponding analysis
of one-dimensional distributions of the constructed 
invariant masses preserves information about interference. 
Such a one-dimensional analysis is simpler than the description of 
the elements of a two-dimensional distribution on the Dalitz plot,
 since a reliable description of the resolution function 
in a multidimensional space is quite a difficult task. 
The two-dimensional approach in describing the distribution
 of events over invariant masses on the Dalitz plot carries more
 information, but the proposed alternative is more efficient
to measure the partial probability of the process $J/\psi \rightarrow
\rho\pi$. This is due to the fact that the main part of the events of 
this process is concentrated in the "corners" of the
three-dimensional  distribution of the cosines of 
the expansion angles of $\pi$ meson pairs. 
Regions where one or another mode of the $J/\psi\rightarrow \rho \pi$  
process predominates can also be distinguished by other conditions 
that include the main set of events of each considered mode. 
However, this does not affect the nature of the distributions 
over the invariant mass for pairs of pions, 
which will be characterized by a "tail" falling to the right. 
This, ultimately, provides the advantage of such an analysis in
finding  the systematic uncertainties associated with the interference with
resonances lying above the $\rho$ meson.

The construction of Dalitz plots and selection criteria based 
on the restriction of the squares of the invariant masses, 
or the invariant masses themselves, are
familiar tools of modern high-energy physics 
used to analyze the processes leading to three particles in
final state. At the same time, for the analysis of such processes, 
plots and selection criteria based on the values of the cosines of
the expansion angles between the final particles can be considered.
Application of the method described above for determining
the probabilities of various $J/\psi \rightarrow \rho \pi$ decay modes 
will allow us to check more accurately  the conservation 
of isotopic invariance for the process under consideration. 
The proposed method can
be useful in the study of the charge asymmetry in the decays  $J/\psi
\rightarrow \rho \pi$ and the so-called $\rho-\pi$ puzzle
\cite{MRK2:1983}, 
if we carry out a similar analysis of various
$\psi(2S)\rightarrow\rho\pi$ decay modes. 
Also, the described method of analysis can be used in the study of the decays
$\Upsilon(1S),\Upsilon(2S)\rightarrow \pi^+\pi^-\pi^0, \phi
K^+K^-,\omega\pi^+\pi^-,K^{*0}(892)K^-\pi^+$ and 
other processes with similar kinematic features.
\section*{Acknowledgments}
The author is grateful to Andrey Shamov and Vladimir Blinov
for their interest to this work.
\bibliographystyle{unsrt}

\end{document}